\documentclass[a4paper, 11pt]{article}

\usepackage[english]{babel} 						
\usepackage[utf8]{inputenc} 						
\usepackage{amsmath} 								
\usepackage{amsfonts} 								
\usepackage{dsfont} 								
\usepackage{amsthm}									
\usepackage{mathtools}								
\usepackage[authoryear]{natbib}						
\usepackage[scale=0.85]{geometry} 					
\usepackage[hidelinks, colorlinks=true,
			allcolors=blue]{hyperref} 				
\usepackage[ruled,linesnumbered]{algorithm2e}		

\usepackage{setspace}
\numberwithin{equation}{section}

\newcommand{\R}{\mathbb{R}}
\newcommand{\N}{\mathbb{N}}
\newcommand{\K}{\mathcal{K}}
\newcommand{\SK}{\mathcal{S}_{q}(\mathcal{K})}

\begin{document}
	
	
\title{Matrix-free Penalized Spline Smoothing with Multiple Covariates}
\author{Julian Wagner \footnote{Trier University / RTG ALOP, Germany}
	\and Göran Kauermann \footnote{LMU Munich, Germany}
	\and Ralf Münnich \footnote{Trier University, Germany}
}
\maketitle


\begin{abstract}
The paper motivates high dimensional smoothing with penalized splines and its numerical calculation in an efficient way.
If smoothing is carried out over three or more covariates the classical tensor product spline bases explode in their dimension bringing the estimation to its numerical limits.
A recent approach by \cite{Siebenborn2019} circumvents storage expensive implementations by proposing matrix-free calculations which allows to smooth over several covariates.
We extend their approach here by linking penalized smoothing and its Bayesian formulation as mixed model which provides a matrix-free calculation of the smoothing parameter to avoid the use of high-computational cross validation.
Further, we show how to extend the ideas towards generalized regression models.
The extended approach is applied to remote sensing satellite data in combination with spatial smoothing.
\end{abstract}




\section{Introduction}
\label{sec:Introduction}


Penalized spline smoothing traces back to \cite{OSullivan1986} but was made popular by \cite{Eilers1996}. The general idea is to replace a smooth unknown function by a high dimensional B-spline basis, where a penalty is imposed on the spline coefficients to guarantee smoothness. The penalty itself is steered by a penalty parameter which controls the amount of penalization. \cite{Wand2003} exhibited the connection between penalized spline smoothing and mixed models and showed how to use mixed model software to estimate the penalty (or regularization) parameter using Maximum Likelihood theory (see also \citealp{Green1993}, \citealp{Brumback1998} or \citealp{Verbyla1999} for earlier references in this line). The general idea is to comprehend the penalty as a prior normal distribution imposed on the spline coefficients. Now, the penalty parameter becomes the (reciprocal of the) prior variance of the random spline coefficients. This link has opened an avenue of flexible smoothing techniques which were proposed in \cite{Ruppert2003}. Penalized spline smoothing achieved general recognition and the method was extended and applied in many areas, as nicely demonstrated in the survey article by \cite{Ruppert2009}. \cite{Kauermann2009} provide a theoretical justification for estimating the smoothing parameter based on mixed model technology. Further details are found in the comprehensive monograph of \cite{Wood2017b}.

In times of \emph{Big Data} we obtain larger and larger data bases which allows for more complex modelling. Though the curse of dimensionality remains (see e.g.\ \citealp{Hastie1990}) we are put in the position of fitting high dimensional models to massive data bases. In particular, this allows to include interactive terms in the model, or putting it differently, we can replace (generalized) additive models of the form
\begin{align}
	\label{eq:gam}
	Y = \beta_0 + s_1(x^1) + ... + s_P(x^P) + \varepsilon
\end{align}
by interaction models of the type
\begin{align}
	\label{eq:int}
	Y = \beta_0 + s(x^1,...,x^P) + \varepsilon.
\end{align}
In (\ref{eq:gam}) the functions $s_p(\cdot)$ are univariate smooth functions, normed in some way to achieve identifiability, while $s(\cdot)$ in (\ref{eq:int}) is a smooth function with a $P$ dimensional argument vector. Using the idea of penalized spline smoothing to fit multivariate smooth functions as in model (\ref{eq:int}) is carried out by using a tensor product of univariate B-spline bases and an appropriately chosen penalty matrix. This leads to a different view of the curse of dimensionality, since the resulting tensor product spline basis increases exponentially in $P$. For instance, using 40 univariate B-splines in each dimension leads to a 64,000 dimensional basis matrix for $P=3$ dimensions. \cite{Wood2017} proposes a discretization scheme as introduced in \cite{Lang2014} to estimate such models. \cite{Li2019} extend the results by exploiting the structure of the model matrices using a blockwise Cholesky decomposition.
If the covariates $x_1,...,x_P$ live on a regular lattice, one can rewrite the entire model and make use of array operations, where the numerical effort in fact then grows only linearly in $P$. This has been proposed in \cite{Currie2006}, see also \cite{Eilers2006}. A different idea to circumvent massive dimensional tensor product B-spline bases has been proposed by \cite{Zenger1991} as so called sparse grids, see also \cite{Bungartz2004} or \cite{Kauermann2013}. The idea is to reduce the tensor product spline dimension by taking the hierarchical construction principle of B-splines into account. However, these simplifications work only in case of regular lattice data and not in general, while we tackle the general case here.

A novel strategy to deal with massive dimensional tensor product spline matrices has been proposed in \cite{Siebenborn2019}, see also \cite{Wagner2019} for extensive technical details. The principal idea is to never construct and store the tensor product spline basis (which in many cases is numerically not even feasible) but to exploit the structure of the tensor product and calculate the necessary quantities by using univariate B-spline bases only. The strategy is referred to as \textsl{matrix-free} calculation, since it only involves the calculation and storage of univariate basis matrices but not of the massive dimensional multivariate basis matrices. We extend these results here by linking penalized spline smoothing to mixed models. Hence, we develop a matrix-free calculation of the necessary quantities to apply a (RE)ML based calculation of the smoothing parameter. This in turn allows to generally apply high dimensional smoothing based on tensor-product spline bases including a data driven calculation (estimation) of the smoothing or regularization parameter, respectively.
We extend these results in two directions. First, we make use of the link between penalized splines and mixed models, which allows to estimate the smoothing parameter as a priori variance of the prior imposed on the spline coefficients. The resulting estimation formulae are simple in matrix notation (see e.g. \citealp{Wood2017}), but for high dimensional tensor product splines their calculation becomes numerically infeasible. We show how to  calculate the quantities with alternative algorithms using a matrix-free strategy. This in turn allows to estimate the smoothing parameter without actually even building the high dimensional and numerically demanding (or infeasible) design matrix. The second extension to the results in \cite{Siebenborn2019} is, that we show how to extend the ideas towards generalized regression models, where a weight matrix needs to be considered.

We apply the routine to estimate organic carbon using the LUCAS (Land Use and Coverage Area frame Survey) data set which contains topsoil survey data including the main satellite spectral bands on the sampling points as well as spatial information (see e.g. \citealp{Toth2013} and \citealp{Orgiazzi2017}). The results show, that spline regression models are able to provide highly improved estimates using large scale multi-dimensional data.

In Section \ref{sec:SplineSmoothing}, we review tensor product based mutltivariate penalized spline smoothing. In Section \ref{sec:Algorithms}, we show how fitting can be pursued without storage of the massive dimensional multivariate spline basis using matrix-free calculation. Section \ref{sec:gam} extends the results towards generalized additive regression, i.e.\ in case of multiple smooth functions and non-normal response. In Section \ref{sec:Application}, we apply the method to the LUCAS data estimating organic carbon in a multi-dimensional setting. Finally, Section \ref{sec:Summary} concludes the manuscript.


\section{Tensor Product Penalized Spline Smoothing}
\label{sec:SplineSmoothing}


We start the presentation by simple smoothing of a single though multivariate smooth function.
Let us, therefore, assume the data
\begin{equation} \label{eq:data}
	\lbrace (x_i,y_i) \in \R^P \times \R \ : \ i=1,\ldots,n \rbrace ,
\end{equation}
where the $y_i \in \R$ are observations of a continuous response variable and the $x_i \in \Omega \subset \R^P$ represent the corresponding value of a continuous multivariate covariate.
The set $\Omega$ is an arbitrary but typically rectangular subset of $\R^P$ containing the covariates.
We assume the model
\begin{equation} \label{eq:mod0}
	y_i = s(x_i) + \varepsilon_i, \quad \varepsilon_i \stackrel{\text{ind}}{\sim} \mathcal{N}(0,\sigma_{\varepsilon}^2) , \quad i=1,\ldots,n ,
\end{equation}
where $s(\cdot) \colon \Omega \rightarrow \R$ is a smooth but further unspecified function.
The estimation of $s(\cdot)$ will be carried out with penalized spline smoothing, initially proposed by \cite{Eilers1996}, see also \cite{Ruppert2003} and \cite{Fahrmeir2013}.
The main idea is to model the function $s(\cdot)$ as a spline function with a rich basis, such that the spline function is flexible enough to capture very complex and highly nonlinear data structures and, in order to prevent from overfitting the data, to impose an additional regularization or penalty term.

\subsection{Tensor Product Splines}

In the case of a single covariate, i.e. $P=1$, let $\Omega \coloneqq [a,b]$ be a compact interval partitioned by the $m+2$ knots
\begin{equation*}
	\K \coloneqq \lbrace a=\kappa_0 < \ldots < \kappa_{m+1}=b \rbrace .
\end{equation*}
Let $\mathcal{C}^{q}(\Omega)$ denote the space of $q$-times continuously differentiable functions on $\Omega$ and let $\mathcal{P}_q(\Omega)$ denote the space of polynomials of degree $q$.
We call the function space
\begin{equation*}
	\SK \coloneqq \lbrace s \in \mathcal{C}^{q-1}(\Omega) :  s|_{\left[ \kappa_{j-1},\kappa_{j} \right]} \in \mathcal{P}_q \left( \left[ \kappa_{j-1},\kappa_{j} \right] \right), \ j=1,\ldots,m+1 \rbrace
\end{equation*}
the space of spline functions of degree $q \in \N_0$ with knots $\K$.
It is a finite dimensional linear space of dimension
\begin{equation*}
	J \coloneqq \text{dim} \left( \SK \right) = m+q+1
\end{equation*}
and with
\begin{equation*}
	\lbrace \phi_{j,q} \ : \ j=1,\ldots,J \rbrace
\end{equation*}
we denote a basis of $\SK$.
For numerical applications the B-spline basis \cite[cf.][]{Boor1978}  is frequently used, whereas the truncated power series basis \cite[cf.][]{Schumaker1981} is often used for theoretical investigations.

To extend the concept of splines to multiple covariates, i.e.\ $P \geq 2$, we implement a tensor product approach.
Let therefore $\mathcal{S}_{q_p}(\K_p)$ denote the spline space for the $p$-th covariate, $p=1,\ldots,P$, and let
\begin{equation*}
	\lbrace \phi_{j_p,q_p}^p \ : \ j_p=1,\ldots,J_p \coloneqq m_p+q_p+1 \rbrace
\end{equation*}
denote the related basis.
The tensor product of basis functions decomposes as product in the form
\begin{equation} \label{def:TPSpline}
	\phi_{j,q} \colon \Omega \coloneqq \Omega_1 \times \ldots \times \Omega_P \subseteq \R^P \rightarrow \R , \quad \phi_{j,q}(x) = \prod_{p=1}^{P} \phi^p_{j_p,q_p}(x^p) ,
\end{equation}
where $j \coloneqq (j_1,\ldots,j_P)$ and $q \coloneqq (q_1,\ldots,q_P)$ denote multiindices and $x = (x^1,\ldots,x^P)^T \in \Omega$ denotes an arbitrary P-dimensional vector. 
The space of tensor product splines is then spanned by these tensor product basis functions, i.e.
\begin{equation*}
	\SK \coloneqq \text{span} \lbrace \phi_{j,q} \ : \ 1 \leq j \leq J \coloneqq (J_1, \ldots, J_P) \rbrace .
\end{equation*}
By definition, this is a finite dimensional linear space of dimension
\begin{equation*}
	K \coloneqq \text{dim}(\SK) = \prod_{p=1}^{P} \text{dim}(\mathcal{S}_{q_p}(\K_p))  = \prod_{p=1}^{P} J_p .
\end{equation*}
Note, that we use the same symbols in the univariate and the multivariate context with the difference that for $ P>1$ we apply a multiindex notation.
Every spline $s \in \SK$ has a unique representation in terms of its basis functions and for computational reasons we uniquely identify the set of multiindices $\lbrace j \in \N^P : 1 \leq j \leq J \rbrace$ in descending lexicographical order as $\lbrace 1,\ldots,K \rbrace$ so that
\begin{equation*}
	s = \sum\limits_{1 \leq j \leq J} \alpha_j \phi_{j,q} = \sum\limits_{k=1}^{K} \alpha_k \phi_{k,q}
\end{equation*}
with unique spline coefficients $\alpha_k \in \R$.

\subsection{Penalized Spline Smoothing}

In order to fit a spline function to the given observations \eqref{eq:data} we apply a simple least squares criterion
\begin{equation} \label{eq:LeastSquares}
	\min\limits_{s \in \SK} \sum\limits_{i=1}^{n} \left( s(x_i) - y_i \right)^2 \ \Leftrightarrow \ \min\limits_{\alpha \in \R^K} \Vert \Phi \alpha - y \Vert_2^2,
\end{equation}
where $\Phi \in \R^{n \times K}$ denotes the matrix of spline basis functions evaluated at the covariates, i.e.\ is element-wise given as
\begin{equation*}
	\Phi[i,k] = \phi_{k,q}(x_i) ,
\end{equation*}
$y \in \R^n$ denotes the vector of the response values, and $\alpha \in \R^K$ denotes the unknown vector of spline coefficients.
To prevent overfitting of the resulting least-squares spline a regularization term
\begin{equation*}
	\mathcal{R} \colon \SK \rightarrow \R_+
\end{equation*}
is included to avoid wiggling behavior of the spline function. This leads to the regularized least squares problem
\begin{equation} \label{eq:RegLeastSquares}
	\min\limits_{s \in \SK} \sum\limits_{i=1}^{n} \left( s(x_i) - y_i \right)^2 + \frac{\lambda}{2} \mathcal{R}(s) ,
\end{equation}
where the regularization parameter $\lambda >0$ controls for the influence of the regularization term.
If $P=1$ and if the B-spline basis is based on equally spaced knots, \cite{Eilers1996} suggest to base the penalty on higher-order differences of the coefficients of adjacent B-splines, i.e.
\begin{equation*}
	\sum\limits_{j=r+1}^{J} \Delta_r(\alpha_j)  = \alpha^T (\Delta_r)^T \Delta_r \alpha = \Vert \Delta_{r} \alpha \Vert_2^2 ,
\end{equation*}
where $\Delta_r(\cdot)$ denotes the $r$-th order backwards difference operator and
$ \Delta_r \in \R^{(J-r) \times J} $
denotes the related difference matrix.
According to \cite{Fahrmeir2013}, this difference penalty is extend to tensor product spline functions by
\begin{equation} \label{eq:DiffPenalty}
	\mathcal{R}_{\text{diff}}(s) \coloneqq \sum\limits_{p=1}^{P}  \alpha^T \left( I_{J_1} \otimes \ldots \otimes I_{J_{p-1}} \otimes \left(\Delta_{r_p}^p \right)^T \Delta_{r_p}^{p} \otimes I_{J_{p+1}} \otimes \ldots \otimes I_{J_P} \right) \alpha ,
\end{equation}
where $\Delta_{r_p}^p \in \R^{(J_p-r_p) \times J_p}$ denotes the $r_p$-th order difference matrix for the $p$-th covariate, $I$ denotes the identity matrix of the respective dimension, and $\otimes$ denotes the Kronecker product.
A more general regularization term for univariate splines which is applicable for arbitrary knots and basis functions is due to \cite{OSullivan1986} and is extend to multivariate splines by \cite{Eubank1988}, see also \cite{Green1993}.
This so called curvature penalty is given as integrated square of the sum of all partial derivatives of total order two, that is
\begin{equation*}
	\mathcal{R}_{\text{curv}}(s) \coloneqq \int\limits_{\Omega} \sum\limits_{p_1=1}^{P} \sum\limits_{p_2=1}^{P} \left( \dfrac{\partial^2}{\partial x_{p_1} \partial x_{p_2}} s(x) \right)^2 \mathrm{d}x .
\end{equation*}
Note that, according to \cite{Siebenborn2019}, it is
\begin{equation} \label{eq:CurvPenalty}
	\mathcal{R}_{\text{curv}}(s) = \sum\limits_{\vert r \vert = 2} \frac{2}{r!} \alpha^T \Psi_r \alpha ,
\end{equation}
where $r \in \N_0^P$ denotes a multiindex and $\Psi_{r} \in \R^{K \times K}$ is element-wise given as
\begin{equation*}
	\Psi_{r}[k,\ell] =  \int\limits_{\Omega} \partial^{r} \phi_{k,q}(x) \partial^{r} \phi_{\ell,q}(x) \mathrm{d}x = \langle \partial^{r} \phi_{k,q}, \partial^{r} \phi_{\ell,q} \rangle_{L^2(\Omega)} .
\end{equation*}
The regularized least squares problem \eqref{eq:RegLeastSquares} in a more convenient form reads
\begin{equation} \label{eq:RegLeastSquaresMatrix}
	\min\limits_{\alpha \in \R^K} \Vert \Phi\alpha - y \Vert_2^2 + \frac{\lambda}{2} \alpha^T \Lambda \alpha
\end{equation}
with $\Lambda \in \R^{ K \times K}$ being an adequate symmetric and positive semidefinite penalty matrix representing the regularization term.
A solution of \eqref{eq:RegLeastSquaresMatrix} is given by a solution of the linear system
\begin{equation} \label{eq:LinearSystem}
	\left( \Phi^T \Phi + \lambda \Lambda \right) \alpha \stackrel{!}{=} \Phi^T y
\end{equation}
with symmetric and positive semidefinite coefficient matrix $\Phi^T \Phi + \lambda \Lambda$.
In the following, we assume that this coefficient matrix is even positive definite, which holds true under very mild conditions on the covariates $x_i$, $i=1,\ldots,n$, and is generally fulfilled in practice \cite[cf.][]{Wagner2019}.
This especially yields that the solution of \eqref{eq:RegLeastSquaresMatrix} uniquely exists and is analytically given as
\begin{equation} \label{eq:RegLeastSMsol}
	\widehat{\alpha} \coloneqq \left( \Phi^T \Phi + \lambda \Lambda \right)^{-1} \Phi^T y .
\end{equation}
Though formula \eqref{eq:RegLeastSMsol} looks simple, for P in the order of three or higher we obtain a numerically infeasible problem due to an exponential growth of the problem dimension $K$ within the number of covariates $P$, i.e.\ $K= \mathcal{O}(2^P)$.
For example for $P=3$ already the storage of the penalty matrix $\Lambda$ requires approximately two gigabyte (GB) of random access memory (RAM) even with using a sparse column format.
Since not only $\Lambda$ has to be stored and since the matrices further need to be manipulated, i.e.\ the the linear system \eqref{eq:LinearSystem} has to be solved, this clearly exceeds the internal memory of common computer systems.
More detail is given in Section \ref{sec:Algorithms}.

\subsection{Regularization Parameter Selection}
\label{subsec:SmoothingParameter}

Selecting the regularization parameter is an important part since it regulates the influence of the regularization term and therefore the amount of smoothness of the resulting P-spline.
We follow the mixed model approach of \cite{Ruppert2003}, see also \cite{Wood2017}. To do so, we assume the prior
\begin{equation}
\label{eq:prior}
\alpha \sim N(0, \sigma_{\alpha}^2 \Lambda^{-})
\end{equation}
where $\Lambda^{-}$ denotes the generalized inverse of $\Lambda$. Given $\alpha$ we assume normality so that
\begin{equation}
\label{eq:density}
y|\alpha \sim N(\Phi \alpha, \sigma_{\varepsilon}^2 I).
\end{equation}
Applying (\ref{eq:prior}) to (\ref{eq:density}) leads to the observation model
$$ y \sim N(0, \sigma_{\varepsilon}^2 (I + \lambda^{-1} \Phi^T \Lambda^{-} \Phi)), $$
where
\begin{equation}
	\label{eq:regpar}
	\lambda = \frac{\sigma_{\varepsilon}^2}{\sigma_{\alpha}^2}
\end{equation}
with
\begin{equation*}
	\sigma_{\varepsilon}^2 = \frac{\Vert \Phi\alpha-y \Vert_2^2}{n}
	\quad \text{ and } \quad
	\sigma_{\alpha}^2 = \frac{\alpha^T \Lambda \alpha}{\text{trace}\left( ( \Phi^T\Phi + \lambda \Lambda )^{-1}\Phi^T\Phi \right)} \quad.
\end{equation*}
Since the variances depend on the parameter $\lambda$ and vice versa, an analytical solution can not be achieved.
However, following \cite{Wand2003} or \cite{Kauermann2005} we can estimate the variances iteratively as follows.
Let $\hat{\alpha}^{(t)}$ be the penalized least squares estimate of $\alpha$ with $\lambda$ being set to $\hat{\lambda}^{(t)}$, then
\begin{equation}
	\left( \hat{\sigma}_{\varepsilon}^2 \right)^{(t)} = \frac{\Vert \Phi\hat{\alpha}^{(t)}-y \Vert_2^2}{n}
	\quad , \quad
	\left( \hat{\sigma}_{\alpha}^2 \right)^{(t)} = \frac{\left(\hat{\alpha}^{(t)}\right)^T \Lambda \hat{\alpha}^{(t)}}{\text{trace}\left( ( \Phi^T\Phi + \hat{\lambda}^{(t)} \Lambda )^{-1}\Phi^T\Phi \right)} \quad ,
\end{equation}
which leads to a fixed-point iteration according to Algorithm \ref{Algorithm:FixedPoint}.
\begin{algorithm}[htb]
	\caption{Fixed point iteration for $\alpha$ and $\lambda$}
	\label{Algorithm:FixedPoint}
	\KwIn{$\hat{\lambda}^{(0)} > 0$}
	\For{$t=0,1,2,\ldots$}{
		$\hat{\alpha}^{(t)} \gets \left( \Phi^T \Phi + \hat{\lambda}^{(t)} \Lambda \right)^{-1} \Phi^T y$ \\
		$\left( \hat{\sigma}_{\varepsilon}^2 \right)^{(t)} \gets \dfrac{\Vert \Phi\hat{\alpha}^{(t)}-y \Vert_2^2}{n} $ \\
		$\left( \hat{\sigma}_{\alpha}^2 \right)^{(t)} \gets \dfrac{\left(\hat{\alpha}^{(t)}\right)^T \Lambda \hat{\alpha}^{(t)}}{\text{trace}\left( ( \Phi^T\Phi + \hat{\lambda}^{(t)} \Lambda )^{-1}\Phi^T\Phi \right)} $ \\
		$\hat{\lambda}^{(t+1)} \gets \dfrac{\left( \hat{\sigma}_{\varepsilon}^2 \right)^{(t)}}{\left( \hat{\sigma}_{\alpha}^2 \right)^{(t)}}$ \\
		\If{$ \vert \hat{\lambda}^{(t+1)} - \hat{\lambda}^{(t)} \vert \leq \mathtt{tol} $}{
			\text{stop}
		}
	}
	$\hat{\alpha}^{(t+1)} \gets \left( \Phi^T \Phi + \hat{\lambda}^{(t+1)} \Lambda \right)^{-1} \Phi^T y$ \\
	\Return $\hat{\alpha}^{(t+1)}$, $\hat{\lambda}^{(t+1)}$
\end{algorithm}

\section{Matrix-free Algorithms for Penalized Spline Smoothing}
\label{sec:Algorithms}


The main task within the P-spline method is (repetitively) solving a linear system system of the form
\begin{equation}
	\label{eq:mainsystem}
	(\Phi^T \Phi + \lambda \Lambda) \alpha \stackrel{!}{=} \Phi^Ty
\end{equation}
for the unknown spline coefficients $\alpha \in \R^K$.
Since the coefficient matrix is symmetric and positive definite, a variety of solution methods such as the conjugate gradient (CG) method of \cite{Hestenes1952} can be applied.
The difficulty is hidden in the problem dimension
\begin{equation*}
	K = \prod_{p=1}^{P} J_p = \mathcal{O}(2^P)
\end{equation*}
that depends exponentially on the number of utilized covariates $P$.
This fact, known as the curse of dimensionality \cite[cf.][]{Bellman1957}, leads to a tremendous growth of memory requirements to store the coefficient matrix with increasing $P$.
Even for a moderate number of covariates $P \geq 3$ the memory requirements for storing and manipulating the coefficient matrix exceed the working memory of customary computing systems.
To overcome this issue and to make the P-spline method applicable also for covariate numbers $P \geq 3$, we present a matrix-free algorithm to solve the linear system \eqref{eq:mainsystem} as well as to estimate the related regularization parameter \eqref{eq:regpar} that require a negligible amount of storage space.

\subsection{Matrix Structures and Operations}
\label{subsec:MatrixStructure}

The basic idea of matrix-free methods for solving a linear system of equations is not to store the coefficient matrix explicitly, but only accesses the matrix by evaluating matrix-vector products.
Many iterative methods, such as the CG method, allow for such a matrix-free implementation and we now focus on computing matrix-vector products with the coefficient matrix
\begin{equation*}
	\Phi^T \Phi + \lambda \Lambda
\end{equation*}
without assembling and storing the matrices $\Phi$ and $\Lambda$.

We first focus on the spline basis matrix $\Phi \in \R^{n \times K}$.
By definition of the tensor product spline space its basis functions are given as point-wise product of the univariable basis functions \eqref{def:TPSpline} and therefore we conclude for the $i$-th column of $\Phi^T$ that
\begin{equation*}
	\Phi^T[,i] =
	\begin{pmatrix}
	\phi_{1,q}(x_i) \\ \vdots \\ \phi_{K,q}(x_i)
	\end{pmatrix} =
	\bigotimes\limits_{p=1}^{P}
	\begin{pmatrix}
	\phi^p_{1,q_p}(x_i^p) \\ \vdots \\ \phi^p_{J_p,q_p}(x_i^p)
	\end{pmatrix} .
\end{equation*}
Defining $\Phi_p \in \R^{n\times J_p}$ as the matrix of the univariate spline basis functions evaluated at the respective covariates, i.e.
\begin{equation*}
	\Phi_p[i,j_p] \coloneqq \phi^p_{j_p,q_p}(x_i^p) , \quad p=1,\ldots,P,
\end{equation*}
it follows
\begin{equation}
	\label{eq:columnexpression}
	\Phi^T[,i] = \bigotimes\limits_{p=1}^{P} \Phi_p^T[\cdot,i] = \left(\bigotimes\limits_{p=1}^{P} \Phi_p[i,\cdot]\right)^T .
\end{equation}
Note that therefore
\begin{equation*}
	\Phi^T =
	\begin{bmatrix} \left(\bigotimes\limits_{p=1}^{P} \Phi_p[1,\cdot]\right)^T , \ldots , \left(\bigotimes\limits_{p=1}^{P} \Phi_p[n,\cdot]\right)^T  \end{bmatrix} \in \R^{K \times n} ,
\end{equation*}
where $\Phi_p[i,\cdot]$ denotes the $i$-th row of $\Phi_p$.
This can be expressed in more compact form as
\begin{equation*}
	\Phi^T
	= \bigodot\limits_{p=1}^P \Phi_p^T ,
\end{equation*}
where $\odot$ denotes the Khatri-Rao product, which is defined as column-wise Kronecker product for matrices with the same number of columns.
Using \eqref{eq:columnexpression} it holds for arbitrary $y \in \R^n$ that
\begin{equation}
	\label{eq:MVPPhi_T}
	\Phi^T y
	= \sum\limits_{i=1}^{n}y[i] v_i ,
\end{equation}
where
\begin{equation}
	v_i \coloneqq \left(\bigotimes\limits_{p=1}^{P} \Phi_p[i,\cdot]\right)^T \in \R^K
\end{equation}
denotes the $i$-th column of $\Phi^T$ and $y[i]$ the $i$-th element of th vector y.
Exploiting this structure, Algorithm \ref{Algorithm:MVPPhi_T} computes the matrix-vector product $\Phi^T y$ by only accessing and storing the small univariate basis matrices $\Phi_p \in \R^{n \times J_p}$, $p=1,\ldots,P$.
\begin{algorithm}[htb]
	\caption{Matrix-vector product with $\Phi^T$}
	\label{Algorithm:MVPPhi_T}
	\KwIn{$\Phi_1,\ldots,\Phi_P,y$}
	\KwOut{$\Phi^T y$}
	$w \gets 0$ \\
	\For{$i=1,\ldots,n$}{
		$v \gets \Phi_1[i,\cdot] \otimes \ldots \otimes \Phi_P[i,\cdot]$ \\
		$w \gets w+y[i]v$
	}
	\Return $w$
\end{algorithm}

In analogy to \eqref{eq:MVPPhi_T}, it holds for arbitrary $\alpha \in \R^K$ that
\begin{equation*}
	\Phi \alpha
	= \left( v_1^T \alpha  , \ldots  , v_n^T \alpha \right)^T
\end{equation*}
such that Algorithm \ref{Algorithm:MVPPhi} computes the matrix-vector product $\Phi \alpha$ again by only accessing and storing the small factors $\Phi_1,\ldots,\Phi_P$.
\begin{algorithm}[htb]
	\caption{Matrix-vector product with $\Phi$}
	\label{Algorithm:MVPPhi}
	\KwIn{$\Phi_1,\ldots,\Phi_P,\alpha$}
	\KwOut{$\Phi \alpha$}
	\For{$i=1,\ldots,n$}{
		$v \gets \Phi_1[i,\cdot] \otimes \ldots \otimes \Phi_P[i,\cdot]$ \\
		$w[i] \gets v^T\alpha$
	}
	\Return $w$
\end{algorithm}

Since for $P>1$ it holds
\begin{equation*}
	J_p \ll K = \prod\limits_{p=1}^{P} J_p ,
\end{equation*}
the storage costs of the univariable basis matrices $\Phi_p$ are negligibly and the Algorithms \ref{Algorithm:MVPPhi_T} and \ref{Algorithm:MVPPhi} allow the computation of matrix vector products with $\Phi$, $\Phi^T$ and $\Phi^T \Phi$ without explicitly forming these infeasibly large matrices.

Regarding the penalty matrix $\Lambda$, the underlying regularization term is crucial.
For the difference penalty \eqref{eq:DiffPenalty}, we note that
\begin{equation} \label{eq:LambdaDiff}
\begin{split}
	\Lambda_{\text{diff}} & = \sum\limits_{p=1}^{P} \left( I_{J_1} \otimes \ldots \otimes I_{J_{p-1}} \otimes \left(\Delta_{r_p}^p \right)^T \Delta_{r_p}^{p} \otimes I_{J_{p+1}} \otimes \ldots \otimes I_{J_P} \right) \\
	& = \sum\limits_{p=1}^{P} \left( I_{L_p} \otimes \Theta_p  \otimes I_{R_p} \right) ,
\end{split}
\end{equation}
where
\begin{equation*}
	L_p \coloneqq \prod_{t=1}^{p-1} J_t , \quad R_p \coloneqq \prod_{t=p+1}^{P} J_t , \quad \Theta_p \coloneqq \left(\Delta_{r_p}^p \right)^T \Delta_{r_p}^{p} \in \R^{J_p \times J_p} .
\end{equation*}
A matrix of the form
\begin{equation*}
	I_{L} \otimes A \otimes I_R
\end{equation*}
with $L,R \in \N$ and quadratic $A \in \R^{J \times J}$ is called normal factor, such that $\Lambda_{\text{diff}}$ is given as a sum of normal factors.
Modifying the idea of \cite{Benoit2001}, multiplication with a normal factor is achieved as follows.
For arbitrary $\alpha \in \R^{LJR}$ and
\begin{equation*}
	B := A \otimes I_{R} \in \R^{JR \times JR}
\end{equation*}
it holds
\begin{equation*}
	(I_{L} \otimes A \otimes I_R) \alpha =
	\left( I_{L} \otimes B \right) \alpha .
\end{equation*}
The matrix $I_{L} \otimes B $ is a block-diagonal matrix consisting of $L$ blocks containing the matrix $B$ in each block.
According to $B$, we decompose the vector $\alpha$ into $L$ chunks $\alpha_{(1)}, \ldots, \alpha_{(L)}$, each of length $JR$ such that
\begin{equation*}
	(I_{L} \otimes A \otimes I_R) \alpha =
	\begin{pmatrix}
		B\alpha_{(1)} \\ \vdots \\ B\alpha_{(L)}
	\end{pmatrix}
\end{equation*}
In order to compute $B\alpha_{(l)}$ for $l=1,\ldots,L$ note that
\begin{equation*}
	B =
	A \otimes I_{R} =
	\begin{bmatrix}
	a_{1,1}I_R & \ldots & a_{1,J}I_R \\
	\vdots & \ddots & \vdots \\
	a_{J,1}I_R & \ldots & a_{J,J}I_R
	\end{bmatrix}
\end{equation*}
such that each row of $B$ consist of the elements of a row of $A$ at distance $R$ apart and zeros for the rest.
Computation of the $t$-th element of $B\alpha_{(l)}$ therefore boils down to the repeated extraction of components of $\alpha_{(l)}$, at distance $R$ apart and starting with element number $t$, forming a vector $z_{l,t}^{\text{in}} \in \R^{J}$, and then multiplying the $t$-th row of $A$ with $z_{l,t}^{\text{in}}$.
The multiplication
\begin{equation*}
	z_{l,t}^{\text{out}} := Az_{l,t}^{\text{in}} \in \R^J \quad,
\end{equation*}
therefore, provides several elements of the matrix-vector product $(I_{L} \otimes A \otimes I_R) \alpha$, where the positions are at distance $R$ apart.
Finally, looping over all $L$ chunks and all $R$ blocks yields Algorithm \ref{Algorithm:MVPNormalFactor} to compute a matrix-vector product of a normal factor $I_{L} \otimes A \otimes I_R$ with an arbitrary vector $\alpha \in \R^{LJR}$ by only storing the matrix $A \in \R^{J \times J}$.
The application of Algorithm \ref{Algorithm:MVPNormalFactor} to each addend in \eqref{eq:LambdaDiff} finally allows to compute matrix-vector products by only forming and storing the small matrices $\Theta_p \in \R^{J_p \times J_p}$, $p=1,\ldots,P$.
\begin{algorithm}[htb]
	\DontPrintSemicolon
	\caption{Matrix-vector product with a normal factor}
	\label{Algorithm:MVPNormalFactor}
	\KwIn{$L,R \in \N$, $A \in \R^{J \times J}$, $\alpha \in \R^{LJR}$}
	\KwOut{$\left( I_{L} \otimes A \otimes I_{R} \right) \alpha \in \R^{LJR}$}
	$\text{base} \gets 0$ \\
	\For(\tcp*[f]{loop over all $L$ chunks}){$l=1,\ldots,L$}{
		\For(\tcp*[f]{loop over all $R$ blocks}){$r=1,\ldots,R$}{
			$\text{index} \gets \text{base}+r$ \\
			\For(\tcp*[f]{form the vector $z_{l}^{\text{in}}$}){$j=1,\ldots,J$}{
				$z_{\text{in}}[j] \gets \alpha[\text{index}]$ \\
				$\text{index} \gets \text{index}+R$
			}
			$z_{\text{out}} \gets A z_{\text{in}}$ \tcp*{compute $z_{l}^{\text{out}}$}
			$\text{index} \gets \text{base}+r$ \\
			\For{$j=1,\ldots,J$}{
				$\alpha[\text{index}] \gets z_{\text{out}}[j]$ \tcp*{store results at the right position}
				$\text{index} \gets \text{index}+R$
			}
		}
		$\text{base} \gets \text{base}+RJ$ \tcp*{jump to the next chunk}
	}
	\Return $\alpha$
\end{algorithm}

If the curvature penalty \eqref{eq:CurvPenalty} is used instead of the difference penalty \eqref{eq:DiffPenalty}, i.e.
\begin{equation*}
	\Lambda_{\text{curv}} = \sum\limits_{\vert r \vert = 2} \frac{2}{r!} \Psi_r ,
\end{equation*}
\cite{Wagner2019} showed that
\begin{equation*}
	\Psi_r = \bigotimes\limits_{p=1}^{P} \Psi_{r_p}^p = \prod\limits_{p=1}^{P} \left( I_{L_p} \otimes \Psi_{r_p}^p \otimes I_{R_p} \right) ,
\end{equation*}
where $\Psi_{r_p}^p$ is the onedimensional counterpart of $\Psi_r$ in the $p$-th direction, i.e.
\begin{equation*}
	\Psi_{r_p}^{p} \in \R^{J_p \times J_p} , \quad \Psi_{r_p}^{p}[j_p,\ell_p]= \left\langle \partial^{r_p} \phi_{j_p,q_p}^p , \partial^{r_p} \phi_{\ell_p,q_p}^p \right\rangle_{L^2(\Omega_p)} , \quad p=1,\ldots,P .
\end{equation*}
Therefore, a matrix vector product with $\Lambda_{\text{curv}}$ can also be computed by means of Algorithm \ref{Algorithm:MVPNormalFactor} by looping backwards over all $P$ normal factors.
That is, setting $w_P := \alpha$, we compute
\begin{equation*}
	w_{p-1} :=  \left( I_{L_p} \otimes \Psi_{r_p}^p \otimes I_{R_p} \right) w_{p}, \quad p=P,\ldots,1
\end{equation*}
using Algorithm \ref{Algorithm:MVPNormalFactor} and obtain $w_0 = \Psi_r \alpha$ as the desired matrix-vector product, again by only storing the small matrices $\Psi_{r_p}^p \in \R^{J_p \times J_p}$, $p=1,\ldots,P$.

\subsection{Matrix-free CG Method}

We now turn back to the main problem of solving the linear system \eqref{eq:mainsystem} for the spline coefficients $\alpha \in \R^K$, where the coefficient matrix
\begin{align*}
	\Phi^T \Phi + \lambda \Lambda \in \R^{K \times K}
\end{align*}
does not fit into the working memory of customary computational systems.
Since the coefficient matrix is symmetric and positive definite, the conjugate gradient method is an appropriate solution method.
A major advantage of the CG method is that the coefficient matrix does not have to be known explicitly but only implicitly by actions on vectors, i.e.\ only matrix-vector products with the coefficient matrix are required.
The CG method can therefore be implemented in a matrix-free manner, in the sense that the coefficient matrix is not required to exist.
Using the algorithms implemented in Section \ref{subsec:MatrixStructure}, matrix-vector products with all components $\Phi^T$, $\Phi$, and $\Lambda$ of the coefficient matrix, and therefore with the coefficient matrix itself, can be computed at negligible storage costs.
The resulting matrix-free version of the CG method to solve the linear system \eqref{eq:mainsystem} is given in Algorithm \ref{Algorithm:MatrixFreeCG}.
Note that, for a fixed $P$, Algorithm \ref{Algorithm:MatrixFreeCG} merely requires the storage of the small factor matrices occurring within the Kronecker and Khatri-Rao product matrices.
Therefore, the storage demand depends only linear on $P$, i.e.\ is $\mathcal{O}(P)$, which is an significant improvement compared to $\mathcal{O}(2^P)$ which is required for the naive implementation with full matrices.
\begin{algorithm}[h!]
	\DontPrintSemicolon
	\caption{Matrix-free CG method for $\alpha$}
	\label{Algorithm:MatrixFreeCG}
	\KwOut{$\left( \Phi^T\Phi + \lambda \Lambda \right)^{-1}\Phi^Ty$}
	$\alpha \gets 0$ \\
	$p \gets r \gets \Phi^Ty$ \tcp*{matrix-vector product by Algorithm \ref{Algorithm:MVPPhi_T}}	
	\While{$\Vert r \Vert_2^2 > \mathtt{tol}$}{
		$v \gets \left( \Phi^T\Phi + \lambda \Lambda \right) p$ \tcp*{matrix-vector product by Algorithm \ref{Algorithm:MVPPhi_T}, \ref{Algorithm:MVPPhi}, and \ref{Algorithm:MVPNormalFactor}}
		$w \gets \Vert r \Vert_2^2 / (p^Tv)$ \\
		$\alpha \gets \alpha + w p$ \\
		$\tilde{r}\gets r$ \\
		$r \gets r - w v$ \\
		$p \gets r + (\Vert r \Vert_2^2 / \Vert \tilde{r} \Vert_2^2) p$ \\
	}
	\Return $\alpha$
\end{algorithm}

It is well known that the CG method finds the exact solution in at most $K$ iterations.
However, since $K$ might be very large, the CG method is in general used as an iterative method in practice.
The rate of convergence of the CG method strongly depends on the condition number of the coefficient matrix which is assumed to be large due to the construction of the linear system \eqref{eq:mainsystem} via a normal equation.
Further, for an increasing number $P$ of covariates, the condition number deteriorates since the number $n$ of observations does not increase to the same extend.
In order to significantly reduce the runtime of Algorithm \ref{Algorithm:MatrixFreeCG}, preconditioning techniques are widely used.
However, since the CG method is implemented matrix-free, the preconditionier has to be matrix-free itself, such that traditional methods cancel out.
According to \cite{Siebenborn2019} a multigrid preconditioner can be utilized provided the curvature penalty with equally spaced knots and B-spline basis is used.
The presented matrix-free CG method, on the contrary, is applicable for arbitrary choices of knots, penalties, and basis functions.
In this case, we can at least implement a simple diagonal preconditioner, that is we choose the diagonal of the coefficient matrix as preconditioner.
To compute this diagonal, again, the coefficient matrix is not allowed to exist.
Since the diagonal of a Kronecker-matrix is simply the Kronecker-product of the diagonals of the factors it remains to compute the diagonal of $\Phi^T\Phi$.
Let $e_k$ denote the $k$-th unit vector such the the $k$-th diagonal element of $\Phi^T\Phi$ reads
\begin{equation*}
	\Phi^T\Phi [k,k] =
	e_k^T \Phi^T\Phi e_k =
	\Vert \Phi e_k \Vert_2^2 =
	\Vert \begin{pmatrix}
	v_1^Te_k \\ \ldots \\ v_n^Te_k
	\end{pmatrix} \Vert_2^2 =
	\Vert \begin{pmatrix}
	v_1^T[k] \\ \ldots \\ v_n^T[k]
	\end{pmatrix} \Vert_2^2 =
	\sum\limits_{i=1}^{n} v_i[k]^2 ,
\end{equation*}
where $v_i \in \R^K$ is defined as in \eqref{eq:MVPPhi_T}.
In summary, the diagonal of the coefficient matrix $\Phi^T \Phi + \lambda \Lambda$ can be computed without explicitly forming the matrix.

\subsection{Matrix-free Regularization Parameter Estimation}
\label{subsec:MFRegPar}

As shown in Section \ref{subsec:SmoothingParameter}, the regularization parameter $\lambda$ can be estimated by means of the fixed-point iteration presented in Algorithm \ref{Algorithm:FixedPoint}.
There, in each iteration step the linear system
\begin{equation*}
	\left( \Phi^T \Phi + \hat{\lambda}^{(t)} \Lambda \right) \alpha \stackrel{!}{=} \Phi^Ty
\end{equation*}
has to be solved, which can now be achieved by Algorithm \ref{Algorithm:MatrixFreeCG}.
It remains to compute the trace of the matrix
\begin{equation*}
	B_t := \left( \Phi^T\Phi + \hat{\lambda}^{(t)} \Lambda \right)^{-1}\Phi^T\Phi \in \R^{K \times K}
\end{equation*}
which is clearly not trivial since $\left( \Phi^T\Phi + \hat{\lambda}^{(t)} \Lambda \right)^{-1}$ can not be computed.
Since $B_t$ is symmetric and positive definite, we follow the suggestions of \cite{Avron2011} who proposed several algorithms to estimate the trace of an implicit, symmetric, and positive semidefinite matrix.
The basic approach, which is due to \cite{Hutchinson1989}, is to estimate the trace of $B_t$ as
\begin{equation*}
	\text{trace}(B_t) \approx \frac{1}{M} \sum\limits_{m=1}^{M} z_m^T B_t z_m ,
\end{equation*}
where the $z_m \in \R^K$ are $M$ independent random vectors whose entries are i.i.d.\ Rademacher distributed, that is
\begin{equation*}
	\text{prob}(z_m = \pm 1) = 1/2 .
\end{equation*}
The advantage of this approach is that the matrix $B_t$ does not have to be explicitly known, but only the $M$ products $z_m^T (B_t z_m)$ need to be efficiently computed.
For computational reasons, we use the reformulation
\begin{equation*}
	\text{trace} (B_t)
	= \text{trace} \left( \left( \Phi^T\Phi + \hat{\lambda}^{(t)} \Lambda \right)^{-1} \Phi^T\Phi \right)
	= K - \text{trace} \left( \left( \Phi^T\Phi + \hat{\lambda}^{(t)}  \Lambda \right)^{-1} \hat{\lambda}^{(t)} \Lambda \right) .
\end{equation*}
Given the random vectors $z_m$, $m=1,\ldots,M$, we first compute the vectors
\begin{equation*}
	\tilde{z}_m^{(t)} \coloneqq \hat{\lambda}^{(t)} \Lambda z_m
\end{equation*}
by means of Algorithm \ref{Algorithm:MVPNormalFactor} and then the vectors
\begin{equation} \label{eq:barz}
	\bar{z}_m^{(t)} \coloneqq \left( \Phi^T\Phi + \hat{\lambda}^{(t)} \Lambda \right)^{-1} \tilde{z}_m^{(t)}
\end{equation}
by Algorithm \ref{Algorithm:MatrixFreeCG}, i.e.\ as solution of a linear system.
The desired estimate of the trace then finally reads
\begin{equation}
	\label{eq:TraceEst}
	\text{trace} \left( \left( \Phi^T\Phi + \hat{\lambda}^{(t)} \Lambda \right)^{-1} \Phi^T\Phi \right) \approx K - \frac{1}{M} \sum\limits_{m=1}^{M} z_m^T \bar{z}_m^{(t)} .
\end{equation}
Summarizing the results, Algorithm \ref{Algorithm:MatrixFreeRegpar} allows for a simultaneous matrix-free estimation of the regularization parameter $\lambda$ and the spline coefficients $\alpha$.
\begin{algorithm}[htb]
	\DontPrintSemicolon
	\caption{Matrix-free fixed-point iteration for $\alpha$ and $\lambda$}
	\label{Algorithm:MatrixFreeRegpar}
	\KwIn{$z_1, \ldots, z_M$, $\hat{\lambda}^{(0)} > 0$}
	\For{$t=0,1,2,\ldots$}{
		$\hat{\alpha}^{(t)} \gets \left( \Phi^T \Phi + \hat{\lambda}^{(t)} \Lambda \right)^{-1} \Phi^T y$ \tcp*{solve linear system by Algorithm \ref{Algorithm:MatrixFreeCG}}
		$\left( \hat{\sigma}_{\varepsilon}^2 \right)^{(t)} \gets \dfrac{\Vert \Phi\hat{\alpha}^{(t)}-y \Vert_2^2}{n}$ \tcp*{matrix-vector product by Algorithm \ref{Algorithm:MVPPhi}}
		\For{$m=1,\ldots,M$}{
			$\tilde{z}_m^{(t)} \gets \hat{\lambda}^{(t)} \Lambda z_m$ \tcp*{matrix-vector product by Algorithm \ref{Algorithm:MVPNormalFactor}}
			$\bar{z}_m^{(t)} \gets \left( \Phi^T\Phi + \hat{\lambda}^{(t)} \Lambda \right)^{-1} \tilde{z}_m^{(t)}$ \tcp*{solve linear system by Algorithm \ref{Algorithm:MatrixFreeCG}}
		}
		$v^{(t)} \gets K- \dfrac{1}{M} \left( \sum\limits_{m=1}^{M} (z_m^{(t)})^T \bar{z}_m^{(t)} \right)$ \\
		$\left( \hat{\sigma}_{\alpha}^2 \right)^{(t)} \gets \dfrac{\left(\hat{\alpha}^{(t)}\right)^T \Lambda \hat{\alpha}^{(t)}}{v^{(t)}}$ \tcp*{matrix-vector product by Algorithm \ref{Algorithm:MVPNormalFactor}}
		$\hat{\lambda}^{(t+1)} \gets \dfrac{\left( \hat{\sigma}_{\varepsilon}^2 \right)^{(t)}}{\left( \hat{\sigma}_{\alpha}^2 \right)^{(t)}}$ \\
		\If{$ \vert \hat{\lambda}^{(t+1)} - \hat{\lambda}^{(t)} \vert \leq \mathtt{tol} $}{
			\text{stop}
		}	
	}
	$\hat{\alpha}^{(t+1)} \gets \left( \Phi^T \Phi + \hat{\lambda}^{(t)} \Lambda \right)^{-1} \Phi^T y$ \tcp*{solve linear system by Algorithm \ref{Algorithm:MatrixFreeCG}}
	\Return $\hat{\alpha}^{(t+1)}$, $\hat{\lambda}^{(t+1)}$
\end{algorithm}

\begin{figure}
	\centering
	\includegraphics[width=0.8\linewidth]{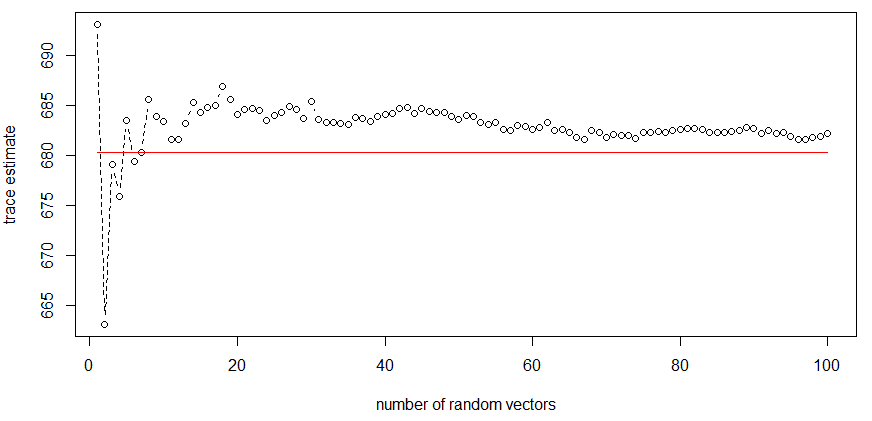}
	\caption{Trace estimates by the Hutchinson estimator versus number $M$ of random vectors.}
	\label{fig:traceest}
\end{figure}

Certainly, the accuracy of the trace estimate \eqref{eq:TraceEst} depends on the number $M$ of utilized random vectors as well as on the particular sample.
For the application we have in mind, however, numerical tests show that already a moderate amount of random vectors yields satisfying estimates.
This is shown in Figure \ref{fig:traceest} where the trace estimates for a simple test problem with $P=2$ covariates and fixed regularization parameter $\lambda$ are graphed versus the number of utilized random vectors.
In this particular example, the true trace (red line) is estimated with sufficient precision already for $M \geq 3$ and the accuracy only slightly improves for an increasing number of random vectors.
More detail and alternative trace estimates are given for example by \cite{Avron2011}.


\section{Generalized Additive Models}
\label{sec:gam}


The above results are given for normal response models. Following the usual generalization from regression to generalized regression we extend the above results now by allowing for an exponential family distribution as well as multiple additive functions.

\subsection{Additive Models}
\label{subsec:Additive}

We discuss first how the above formulas extend to the case of multiple smooth functions. That is we replace (\ref{eq:mod0}) by
\begin{equation*}
	y_i = \sum_{j=1}^{I} s_{(j)}(x_{(j),i}) + \varepsilon_i,
\end{equation*}
where $x_{(j)}$ is a vector of continuous covariates with $x_{(j)} \in \mathbb{R}^{K_{(j)}}$ and $s_{(j)}(\cdot)$ a smooth function, which is estimated by penalized splines. To do so, we replace it by a (Tensor product) matrix $\Phi_{(j)} \in \R^{n \times K_{(j)}}$ constructed as above with corresponding spline coefficient $\alpha_{(j)} \in \R^{K_{(j)}}$. This leads, again, to the massive dimensional spline basis matrix
\begin{equation*}
	\Phi = \left[ \Phi_{(1)}, \ldots, \Phi_{(I)} \right] \in \R^{n \times K}, \quad K := \sum\limits_{j=1}^{I}K_{(j)} .
\end{equation*}
However, since
\begin{equation*}
	\Phi \alpha = \left[ \Phi_{(1)}, \ldots, \Phi_{(I)} \right] \begin{pmatrix}
	\alpha_{(1)} \\ \vdots \\ \alpha_{(I)}
	\end{pmatrix} = \sum\limits_{j=1}^{I} \Phi_{(j)} \alpha_{(j)}
\end{equation*}
and
\begin{equation*}
	\Phi^Ty = \begin{bmatrix}
	\Phi_{(1)}^T \\ \vdots \\ \Phi_{(I)}^T
	\end{bmatrix} y =
	\begin{pmatrix}
	\Phi_{(1)}^Ty \\ \vdots \\ \Phi_{(I)}^Ty ,
	\end{pmatrix} ,
\end{equation*}
the methods presented in Algorithm \ref{Algorithm:MVPPhi_T} and \ref{Algorithm:MVPPhi} are still applicable to compute matrix-vector products with $\Phi$, $\Phi^T$, and $\Phi^T \Phi$.
For the related penalty matrix
\begin{equation*}
	\Lambda(\lambda) := \mbox{ block diag }(\lambda_{(j)}\Lambda_{(j)})
\end{equation*}
the matrix-vector products
\begin{equation*}
	\Lambda(\lambda) \alpha =  \begin{pmatrix}
	\lambda_{(1)}\Lambda_{(1)} \alpha_{(1)} \\ \vdots \\ \lambda_{(I)}\Lambda_{(I)} \alpha_{(I)}
	\end{pmatrix}
\end{equation*}
can also still be computed by Algorithm \ref{Algorithm:MVPNormalFactor}.
Thus, the matrix-free CG method \ref{Algorithm:MatrixFreeCG} is straightforwardly extended to the case of an additive model.

In order to estimate the regularization parameters we assume normality for $\alpha_{(j)}$ as in \eqref{eq:RegLeastSMsol} and mutual independence of $\alpha_{(j)}$ and $\alpha_{(l)}$ for $j \neq l$, so that
\begin{equation*}
	\alpha_{(j)} \sim N(0, \sigma_{(j)}^2 \Lambda_{(j)}^-).
\end{equation*}
We can now estimate the prior variances $\sigma_{(j)}^2$ by extending \eqref{eq:regpar} as follows.
Let $$D_{(j)}^- = \mbox{ block diag}(0,...,0, \Lambda_{(j)}^-,0,...,0),$$ that is $D_{(j)}^-$ is a block diagonal matrix with zero entries except of the j-th block diagonal.
This leads to
\begin{equation*}
	\hat{\sigma}_{(j)}^2 = \frac{  \alpha_{(j)}^T \Lambda_{(j)} \alpha_{(j)}}{ \text{trace}((I + \Phi \Lambda(\lambda)^- \Phi)^{-1} \Phi \lambda_{(j)} D_{(j)}^- \Phi^T) }.
\end{equation*}
The above formula can be approximated through
\begin{equation*}
	\hat{\sigma}_{(j)}^2
	\approx	\frac{  \alpha_{(j)}^T \Lambda_{(j)} \alpha_{(j)}}{ \text{trace}((\Phi_{(j)}^T \Phi_{(j)} + \lambda_{(j)} \Lambda_{(j)})^{-1} \Phi_{(j)}^T \Phi_{(j)}) }
\end{equation*}
which results if we penalize only the $j$-th component.
As in the case of one single smooth function, the variances $\hat{\sigma}_{(j)}^2 $ can be estimated by means of formula \eqref{eq:TraceEst} by only taking the $j$-th component into account.

Finally, since all required components can be computed matrix-free, Algorithm \ref{Algorithm:MatrixFreeRegpar} can directly be extended to the additive model with only slight modifications.

\subsection{Generalized Response}

The results of the previous chapter are readily extended towards generalized regression models. This leads to iterated weighted fitting, that is applying the algorithm of the previous section with weights included in an iterative manner. Following the standard derivations of generalized additive models (see e.g.\ \citealp{Hastie1990} or \citealp{Wood2017b}), we assume the exponential family distribution model
\begin{equation*}
	y|x \sim \exp\{ (y\theta - \kappa(\theta)) \} c(y)
\end{equation*}
where
\begin{equation} \label{eq:link}
	\theta = h(s(x))
\end{equation}
with $h(\cdot)$ as invertible link or response function. The terms $\theta$ and $s(x)$ are linked through the expected value
\begin{equation*}
	E(y|x) = \frac{\partial \kappa(\theta)}{\partial \theta} = \mu(\theta) = \mu(h(s(x))) .
\end{equation*}
Minimization (\ref{eq:LeastSquares}) is then replaced by
\begin{equation*}
	\underset{s\in S_q(\mathcal{K})}{\min} -\sum_{i=1}^n (y_i\theta_i - \kappa(\theta_i)) + \frac{\lambda}{2} R(s).
\end{equation*}
This modifies (\ref{eq:RegLeastSquaresMatrix}) to the iterative version
\begin{equation}\label{eq:iterFisher}
	\alpha^{(t+1)} = \alpha^{(t)} + I(\alpha^{(t)}, \lambda)^{-1} s(\alpha^{(t)}, \lambda)
\end{equation}
where
\begin{equation*}
	I(\alpha^{(t)}, \lambda) = \Phi^T W_2^{(t)} \Phi + \lambda \Lambda
\end{equation*}
is the (penalized) Fischer matrix and
\begin{equation*}
	s(\alpha^{(t)}, \lambda) = \Phi^T W_1^{(t)} \left( y- h(\Phi \alpha^{(t)}) \right) - \lambda \Lambda \alpha^{(t)}
\end{equation*}
is the penalized score.
The weight matrices are constructed from
\begin{align}
\label{eq:WMatrix}
\begin{split}
	&W_1^{(t)} = \text{diag}(h'(\Phi \alpha^{(t)})) \\
	&W_2^{(t)} = \text{diag}(h(\Phi \alpha^{(t)}) \frac{\partial^2 \kappa(\theta^{(t)})}{\partial \theta \partial \theta^T} h(\Phi \alpha^{(t)})^T)
\end{split}
\end{align}
where $h'(\cdot)$ is the first order derivative of $h(\cdot)$ and $\partial^2 \kappa(\theta) /(\partial \theta \partial \theta^T)$ results as the variance contribution in the exponential family distribution.
In case of a canonical link, that is $\theta = \Phi \alpha$, one gets $h(\cdot)=id(\cdot)$ so that $W_1$ equals the identity matrix and $W_2$ is the diagonal matrix of variances.

In order to solve the linear system $$I(\alpha^{(t)}, \lambda)^{-1} s(\alpha^{(t)}, \lambda)$$ we have to slightly modify Algorithm \ref{Algorithm:MatrixFreeCG} since we now need to compute matrix-vector products with $I(\alpha^{(t)}, \lambda)$.
Therefore note that $W_2^{(t)} \in \R^{K \times K}$ is a diagonal matrix and can therefore be stored as a vector $w_2^{(t)} \in \R^K$.
The matrix-vector product is then given as
\begin{equation*}
	I(\alpha^{(t)}, \lambda) \alpha^{(t)}
	= \Phi^T W_2^{(t)} \Phi \alpha^{(t)}  + \lambda \Lambda \alpha^{(t)}
	= \Phi^T \left( w_2^{(t)} * \left( \Phi \alpha^{(t)} \right) \right)  + \lambda \Lambda \alpha^{(t)} ,
\end{equation*}
where $*$ denotes element-wise vector-vector multiplication.
The product can then still be computed by means of the methods presented in Algorithm \ref{Algorithm:MVPPhi}, \ref{Algorithm:MVPPhi_T}, and \ref{Algorithm:MVPNormalFactor}.
Since the same modification with $w_1^{(t)} \in \R^K$ instead of $W_1^{(t)} \in \R^{K \times K}$ is possible to compute the right-hand side vector $s(\alpha^{(t)}, \lambda)$, the matrix-free CG method can be applied to solve the linear system in each iteration.

Note, that from a computational point of view, the major difference to the conventional spline regression model is that the P-spline model is fitted by solving one linear system, whereas the generalization requires solving a linear system (of the same size) per iteration.


In order to estimate the regularization parameter one can proceed in analogy to Section \ref{subsec:SmoothingParameter} and obtain $$\hat{\lambda}^{(t)} = \frac{\left( \hat{\sigma}_{\varepsilon}^2 \right)^{(t)}}{\left( \hat{\sigma}_{\alpha}^2 \right)^{(t)} },$$
where in the generalized model it is
\begin{align*}
	& \left( \hat{\sigma}_{\varepsilon}^2 \right)^{(t)} = \frac{\Vert h'(\Phi\hat{\alpha}^{(t)})-y \Vert_2^2}{n} \\
	& \left( \hat{\sigma}_{\alpha}^2 \right)^{(t)}
	= \frac{\left(\hat{\alpha}^{(t)}\right)^T \Lambda \hat{\alpha}^{(t)}}{\text{trace}\left( ( \Phi^TW_2^{(t)}\Phi + \hat{\lambda}^{(t)} \Lambda )^{-1}\Phi^TW_2^{(t)}\Phi \right)}
	= \frac{\left(\hat{\alpha}^{(t)}\right)^T \Lambda \hat{\alpha}^{(t)}}{K - \text{trace}\left( ( \Phi^TW_2^{(t)}\Phi + \hat{\lambda}^{(t)} \Lambda )^{-1} \lambda^{(t)}\Lambda \right)} .
\end{align*}
Since matrix-vector products with $W_2^{(t)}$ can be performed by element-wise vector-vector products, we can estimate the trace of $$(\Phi^TW_2^{(t)}\Phi + \hat{\lambda}^{(t)} \Lambda )^{-1} \lambda^{(t)}\Lambda$$ in analogy to \eqref{eq:TraceEst} by replacing \eqref{eq:barz} with $$\bar{z}_m^{(t)} :=\left( \Phi^TW_2^{(t)}\Phi + \hat{\lambda}^{(t)} \Lambda \right)^{-1} \tilde{z}_m^{(t)}.$$
Also the extension of the generalized response to an additive model is then straightforward in analogy to Section \ref{subsec:Additive}.

\section{Estimating Organic Carbon using LUCAS Data}
\label{sec:Application}



The presence of organic carbon is used to provide information on soil erosion (cf.\ \citealp{fang2018visible}, p.\ 4).
To estimate organic carbon, in general, regression and calibration methods are applied. Applying non-linear regression methods on large data sets, however, urge the need of introducing efficient algorithms to provide appropriate estimates and predicts.
The use of non-parametric regression models like P-splines, so far, was suffering from high computational efforts.
In the following application, we show that multidimensional penalized spline regression can be applied successfully to estimate organic carbon using soil and satellite data.


In our application, we estimate organic carbon, denoted by $Y$ using auxiliary information from the LUCAS data set. The LUCAS topsoil survey data provide information on the main properties and multispectral reflectance data of topsoil in 23 EU member states based on a grid of observations. The database comprises 19,967 geo-referenced samples. The data set as well as its sampling design and methodology are described in \cite{Toth2013}. In addition to the soil properties, the data set comprises diffuse high-resolution reflectance spectra  for all samples from 0.400 to 2.500 $\mu$m with 0.5 nm spectral resolution. For this study these spectra were resamples according to the spectral characteristics of ESA's Optical high-resolution Sentinel-2 satellite mission (492.4 $\mu$m, 559.8 $\mu$m, 664.6 $\mu$m, 704.1 $\mu$m, 740.5 $\mu$m, 782.8 $\mu$m, 832.8 $\mu$m, 864.7 $\mu$m, 1613.7 $\mu$m, 2202.4 $\mu$m) and used as input variables for the regression model to assess the potential of Sentinel-2 for organic carbon monitoring.

As covariate information $X := [X_1,X_2,X_3]$, we use the first three satellite spectral bands of the visible light since they are known to be appropriate indicators for organic carbon.
Adding further spectral bands or using principal components from the band data did not show an improvement of the results.
Also we make use of spatial information $U=[U_1,U_2]$ given by longitude and latitude, respectively.
A RESET test for a linear model on the utilized variables using the \texttt{resettest($\cdot$)} with standard settings from R's \texttt{lmtest} package \cite[cf.][]{Zeileis2002} suggest to reject the null hypothesis on linearity.

As models, we propose the spline models introduced in this paper using the organic carbon as the dependent variable and either spatial information $U$, satellite spectral information $X$, or both in an additive P-spline model on the sampled locations.
In extension to \cite{Wagner2017}, where negative predicts had to be avoided using non-negativity shape constraints, we suggest to induce positivity by adding an exponential link, i.e.\ we model $Y$ as normally distributed with $$E(Y|x) = \exp(s(x))\quad,$$
according to Section \ref{sec:gam}. This leads to weight matrices
\begin{equation*}
	W_1^{(t)} = \text{diag}(\exp(s(x)))
\end{equation*}
and
\begin{equation*}
	W_2^{(t)} = \text{diag} (2\exp(s(x)))
\end{equation*}
as substitutes for \eqref{eq:WMatrix}.
Additionally, we include the linear model as well as the generalized linear model using both $X$ and $U$, leading to eight different models.

Table \ref{tab:models} provides information on the utilized model, the corresponding residual sum of squares (RSS), the Akaike information criterion (AIC), the runtime in seconds for fitting the model with fixed regularization parameter (run\_single), the runtime in seconds for fitting the model with iteratively updated regularization parameter (run\_total), as well as the availability of negative predicts (neg).

\begin{table}[htb]
\centering
\begin{tabular}{c|l|r|r|r|r|c}
	Nr. &model & RSS & AIC & run\_single & run\_total & neg \\ \hline
	1 & $y = x+u$ & 238.4207 & 246016.80 & -- & $<$1.00 & Y \\
	2 & $y = exp(x+u)$ & 8757.9102 & 383028.70 & -- & $<$1.00 & N \\
	3 & $y = s(u)$ & 341.6895 & 221550.30 & 7.43 & 58.83 & Y \\
	4 & $y = s(x)$ & 86.1090 & 169971.40 & 26.50 & 593.60 & Y \\
	5 & $y = s_1(x) + s_2(u)$ & 67.2516 & 161330.70 & 28.15 & 567.42 & Y \\
	6 & $y = exp(s(u))$ & 345.1930 & 221712.90 & 27.45 & 118.55 & N  \\
	7 & $y = exp(s(x))$ & 89.7715 & 170874.40 & 908.15 & 1604.26 & N \\
	8 & $y = exp(s_1(x)+s_2(u))$ & 61.1155 & 156456.90 & 1071.43 & 1948.86 & N	
\end{tabular}
\caption{Comparison of the utilized models.}
\label{tab:models}
\end{table}

\begin{figure}[htb]
\centering
\includegraphics[width=0.32\linewidth]{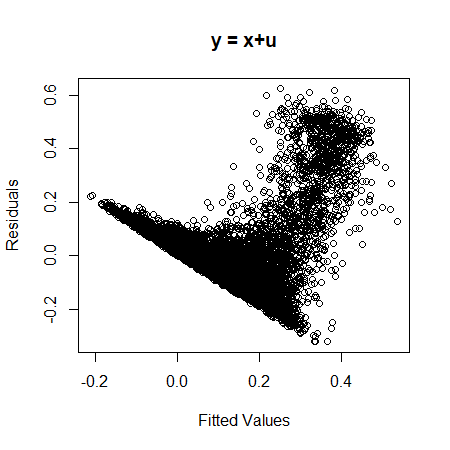}
\includegraphics[width=0.32\linewidth]{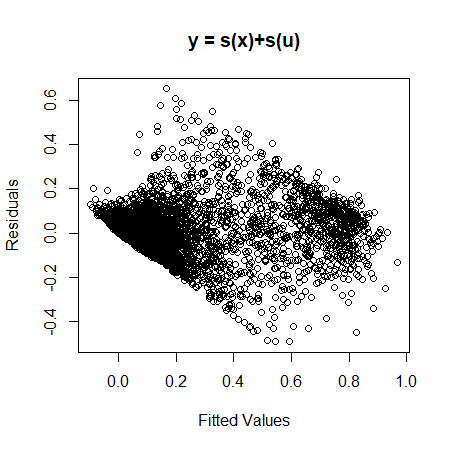}
\includegraphics[width=0.32\linewidth]{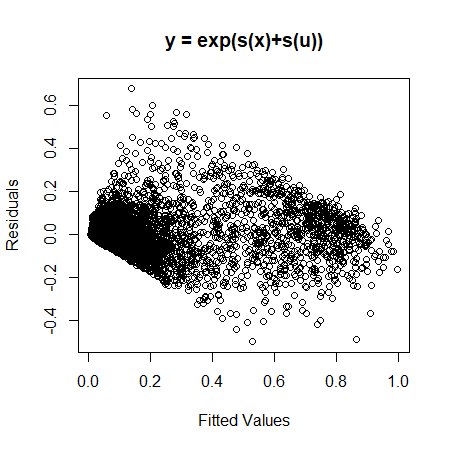}
	\caption{Plots of fitted values versus residuals for the models 1 (left), 5 (middle) and 8 (right).}
	\label{fig:Residuals}
\end{figure}

The results show that linear models provide unacceptable residual sum of squares which is in line with the rejection of the test on linearity. Further, the non-exponential models suffer from negative predicts, which were easily removed using the exponential form. In general, the pure spatial information does not provide high accuracies of the model. However, the combination of spatial and spectral information yields in all cases the best model evaluations. This finally suggest that model 8 is outperforming all other models considerably.
The only drawback is the higher computation time which is induced by applying the exponential form.
These results can also be drawn from the residual plots, which are depicted in Figure \ref{fig:Residuals}.

The selected residual plots are all using spatial and spectral information. Models 1 and 5 show significant patterns which indicate a violation against model assumptions.
Only in the last model 8, the normality assumption is considerably met.  However, some outliers may still leave some space for further finetuning while integrating further soil information or using adequate transformations.

Finally, maps are presented for models 5 and 8. In the left graph, the negative predicts can be seen especially in Eastern Europe. Model 8 provides sensible predicts over Europe.
\begin{figure}[htb]
	\centering
	\includegraphics[width=0.45\linewidth]{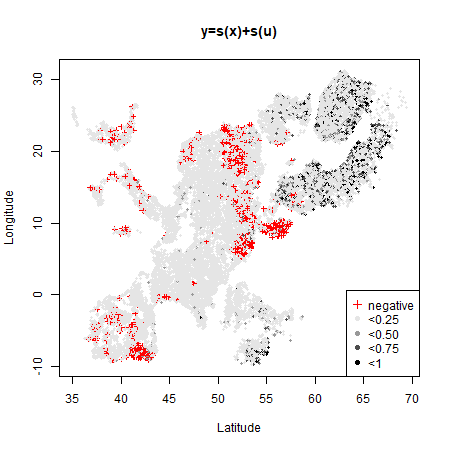}
	\includegraphics[width=0.45\linewidth]{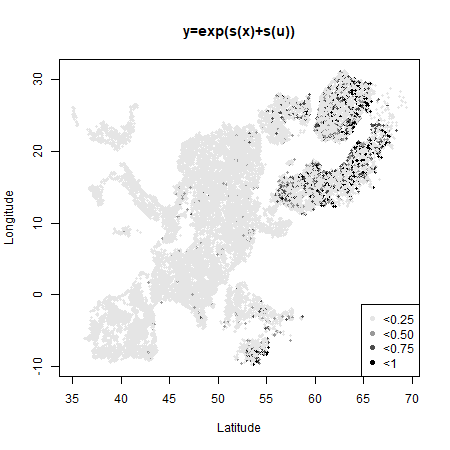}	
\caption{Maps of model predicts from models 5 (left) and 8 (right) for organic carbon on the LUCAS sample}
\label{fig:maps}
\end{figure}


\section{Summary}
\label{sec:Summary}


Penalized spline smoothing is a very powerful and widely used method in non-parametric regression.
In three or more dimensions, however, the method reaches computational limitations due to the curse of dimensionality.
More specifically, storing and manipulating the resulting massive dimensional design matrices rapidly exceed the working memory of customary computer systems.

A recent approach by \cite{Siebenborn2019} circumvents storage expensive implementations by proposing matrix-free calculations, which are briefly presented in this paper. We extended their results by making use of the link between penalized splines and mixed models, which allows to estimate the smoothing parameter and showed how this estimation is performed in a matrix-free manner.
This especially prevents the need of computational expensive cross-validation methods for determining the regularization parameter.
Further, we extended their results towards generalized regression models, where a weight matrix needs to be considered, and showed how these computations can also be performed matrix-free.

An application to estimating organic carbon has shown the straight forward and efficient application of the methods including generalized additive models with an exponential link.


\section*{Acknowledgment}
This work has been supported by the German Research Foundation (DFG) within the research training group ALOP (GRK 2126).
The third author was supported by the RIFOSS project which is funded by the German Federal Statistical Office.


\bibliographystyle{apalike}
\bibliography{citations}


\end{document}